\documentclass[12pt]{article}
\usepackage{cite, bm}
\def\be{\begin{equation}}
\def\ee{\end{equation}}
\def\bea{\begin{eqnarray}}
\def\eea{\end{eqnarray}}

\begin{document}

\begin{center}
{\Large{\bf Wave Propagation in Periodic Plasma Media: A Semiclassical Approach}}

\vskip .5cm {\large Reza Torabi}
\vskip .1cm
{\it Department of Physics and Astronomy, University of Calgary, Calgary, Alberta T2N 1N4, Canada}\\
{\sl e-mail: reza.torabi@ucalgary.ca}\\
\end{center}

\begin{abstract}

The propagation of electromagnetic waves in unmagnetized periodic
plasma media is studied using the semiclassical wave packet
approximation. The formalism gives rise to Berry effect terms in the
equation of motion. The Berry effect manifests itself as Rytov
polarization rotation law and the polarization-dependent Hall
effect. The formalism is also applied to the case of non-periodic
inhomogeneous plasma media.

\end{abstract}

\vskip .5cm
\newpage

\section{Introduction}

The propagation of electromagnetic waves in inhomogeneous plasmas is
an active field of research \cite{Hacquin,Cho,Malik} because most of
the matter in the universe is in the plasma state \cite{Chen} and
real plasmas are inhomogeneous. A periodic plasma medium is an
example of inhomogeneous plasmas. Such a medium can be used to
control and confine light because they are analogous to photonic and
phononic crystals that permit the passage of waves only at certain
frequencies. Ionosphere is another example of inhomogeneous plasma
media that the study of its periodicity is of interest in radio
physics \cite{Bakhmet eva}.

Recently, attention has been paid to the phenomena of polarization
transport in photonic \cite{Bliokh1,Onoda1,Onoda2} and phononic
\cite{Torabi1} crystals. These phenomena are related to the
topological Berry phase \cite{Berry} which is a non-integrable phase
factor arising from the adiabatic transport of a system around a
closed path in the parameter space. In this paper we study the
propagation of electromagnetic waves in an unmagnetized slow varying
periodic plasma medium via the semiclassical wave packet
approximation. The wave packet approximation enables us to solve
problems in the presence of periodic structures. It has been used in
the study of spin transport in solids \cite{Culcer,Chang,Sundaram}
and also in the photonic \cite{Onoda1,Onoda2} and phononic
\cite{Torabi1} crystals. The formalism yields to the rotation of the
polarization plane of a linear polarized wave that is known as Rytov
rotation \cite{Rytov,Vladimirskii}. This is in contrast with the
Faraday rotation which is due to the presence of a magnetic field in
the medium. In addition, the formalism gives rise to a Berry
curvature term in the equation of motion that yields to the spin
(polarization) Hall effect \cite{Torabi2,Dooghin,Liberman}.
According to this effect, waves of left and right circular
polarization deflect perpendicular to the direction of propagation
oppositely.

The paper is organized as follows. In section II, we derive the wave
equation in a periodic plasma medium and introduce Bloch waves as a
complete set to construct the wave packet. Section III is devoted to
the dynamics of the wave packet. Finally, in section IV we apply the
formalism to the case of non-periodic inhomogeneous media. In this
case, Berry curvature has the form of opposite signed magnetic
monopoles located at the origin of the momentum space. The results
for non-periodic inhomogeneous plasma media coincides with the
results of \cite{Torabi3} obtained from the post geometric
approximation.

\section{Wave equation and the Bloch waves}

An Inhomogeneous plasma medium can be generated, for instance, by
applying an external electric field. In this case, the balance
equation of the forces in the steady state is
\[
\rho_{\alpha 0}({\bm x})q_\alpha {\bm E}_0 ({\bm
	x})-m_\alpha \nabla
p_{\alpha0}=0,
\]
where $\rho_{\alpha 0}$ is unperturbed mass density and $\alpha=e,i$
represent electrons and ions. We are going to derive the equation
that determines the wave properties of an inhomogeneous unmagnetized
plasma. In this line, similar to  the derivation of the wave
equation in homogeneous plasmas \cite{Krall}, we consider small
harmonic perturbation about the steady state
\[
\rho_{\alpha}(\bm{x})=\rho_{\alpha0}(\bm{x})+\rho_{\alpha1}(\bm{x})e^{-i\omega
	t},
\]
\[
\bm{V}_{\alpha}=\bm{V}_{\alpha1}(\bm{x})e^{-i\omega t},
\]
\[
\bm{E}={\bm E}_0({\bm x})+\bm{E}_{1}(\bm{x})e^{-i\omega t},
\]
\[
\bm{B}=\bm{B}_{1}(\bm{x})e^{-i\omega t},
\]
but we should notice that the density $\rho_{\alpha 0}(\bm{x})$ is
now a periodic function of $\bm{x}$. Putting the above equations
into the two fluid momentum equation and keep linear terms, yields
to the linearized momentum equation
\[
-i\omega \rho_{\alpha 0}({\bm x}){\bm V}_{\alpha1}({\bm
	x})=\frac{q_\alpha \rho_{\alpha0}({\bm x})}{m_\alpha}{\bm E}_1({\bm
	x})+\rho_{\alpha1}({\bm x})\frac{\nabla p_{\alpha0}}{
	n_{\alpha0}({\bm x})}-\gamma\nabla \bigg{(}\frac{\rho_{\alpha1}({\bm
		x})p_{\alpha0}}{\rho_{\alpha0}({\bm x})} \bigg{)}.
\]
In deriving the above, the balance equation and the relation
$p_{\alpha1}=\gamma p_{\alpha0}\rho_{\alpha 1}({\bm x})e^{-i\omega
	t}/\rho_{\alpha  0}({\bm x})$ have been used. One the other hand,
the linear Maxwell equations are
\[
\nabla \times
B_1=-\frac{i\omega}{c}E_1+\frac{4\pi}{c}\sum_{\alpha=e,i}\rho_{\alpha
	0}({\bm x})q_{\alpha} V_{\alpha 1}({\bm x})
\]
and $\nabla \times E_1=\frac{i\omega}{c}B_1$ which, by taking a curl
from both sides, yields to
\[
\nabla \times \nabla \times E_1=\frac{i\omega}{c}\nabla \times B_1
\]
Combining the linearized momentum equation with linearized Maxwell's
equations, results in the wave equation
\[
\nabla\times\nabla\times
\bm{E}_{1}-k_{0}^{2}\bigg{(}1-\frac{\omega_{pe}^{2}({\bm
		x})}{\omega^{2}}-\frac{\omega_{pi}^{2}({\bm
		x})}{\omega^{2}}\bigg{)}\bm{E}_{1}=\frac{4\pi}{c^2}\sum_\alpha
\frac{q_\alpha}{m_\alpha}\bigg{[}\gamma\nabla
\bigg{(}\frac{\rho_{\alpha 1}({\bm x})p_{\alpha0}}{\rho_{\alpha
		0}({\bm x})}\bigg{)}-\frac{\rho_{\alpha 1}({\bm x})}{\rho_{\alpha
		0}({\bm x})}\nabla p_{\alpha0}\bigg{]},
\]
where $k_{0}=\omega/c$,$c$ is the speed of light and
$\omega_{p\alpha}^{2}({\bm x})=4\pi \rho_{\alpha 0}({\bm
	x})q_{\alpha}^2$ is the plasma frequency for particles of species
$\alpha$. The cold plasma approximation and assuming stationary and
inhomogeneous density is usually used to describe wave propagation
in inhomogeneous media \cite{Ginzburg,Hacquin}. Under the cold
plasma approximation, one can drop the right hand side of the
equation and, thus, the wave equation takes the form
\begin{equation}
\label{1} \nabla\times\nabla\times
\bm{E}_{1}=k_{0}^{2}\bigg{(}1-\frac{\omega_{pe}^{2}({\bm
		x})}{\omega^{2}}-\frac{\omega_{pi}^{2}({\bm
		x})}{\omega^{2}}\bigg{)}\bm{E}_{1}.
\end{equation}
Introducing
\[
\epsilon(\bm{x},\omega)=\bigg{(}1-\frac{\omega_{pe}^{2}({\bm
		x})}{\omega^{2}}-\frac{\omega_{pi}^{2}({\bm
		x})}{\omega^{2}}\bigg{)}\equiv  n^{2}(\bm{x},\omega),
\]
and using, $\nabla\times\nabla\times
\bm{E}_{1}=\nabla(\nabla\cdot{\bf E}_1)-\nabla^2 {\bf E}_1$, to
rewrite (1) as
\begin{equation}
\nabla^2 {\bf E}_1-\nabla(\nabla\cdot{\bf
	E}_1)+\frac{\omega^2}{c^2}n^2({\bm x},\omega){\bf E}_1=0.
\end{equation}
We decompose ${\bf E}_1$ into a solenoidal part, ${\bf E}_{1\bot}$,
and an irrotational part, ${\bf E}_{1\|}$, according to Helmholtz's
decomposition, ${\bf E}_1={\bf E}_{1\bot}+{\bf E}_{1\|}$. The
solenoidal and irrotational parts correspond to transverse and
longitudinal modes of the wave, respectively. Thence, by taking the
curl and the divergence of (2), we obtain two coupled equations
\begin{equation}
\nabla\times\nabla^2{\bf
	E}_{1\bot}+\frac{\omega^2}{c^2}n^2\nabla\times{\bf
	E}_{1\bot}=-\frac{\omega^2}{c^2}[(\nabla n^2)\times {\bf
	E}_{1\bot}+(\nabla n^2)\times {\bf E}_{1\|}],
\end{equation}
and
\begin{equation}
n^2\nabla\cdot{\bf E}_{1\|}=-(\nabla n^2)\cdot{\bf E}_{1\|}-(\nabla
n^2)\cdot{\bf E}_{1\bot}.
\end{equation}

In a smoothly inhomogeneous medium, one can neglect terms on the
right hand sides of (3) and (4) since $\nabla n^2$ is very small. So
we are left with
\begin{equation}
\nabla\times[\nabla^2{\bf E}_{1\bot}+\frac{\omega^2 n^2}{c^2}{\bf
	E}_{1\bot}]=0,
\end{equation}
and
\[
\nabla\cdot {\bf E}_{1\|}=0.
\]
Therefore (3) and (4) decouple in this approximation, which is
called adiabatic approximation. The solenoidal part is divergence
free, $\nabla\cdot {\bf E}_{1\bot}=0$, so one can readily show that
the divergence of the vector in the bracket of (5) is zero, too.
Since, the divergence and the curl of ${\bf E}_{1\bot}$ vanish,
according to the Helmholtz's theorem, we can conclude that this
vector is identically zero, i. e.,
\begin{equation}
\nabla^2{\bf E}_{1\bot}+\frac{\omega^2 n^2}{c^2}{\bf E}_{1\bot}=0.
\end{equation}
Similar argument for the irrotational part shows that ${\bf
	E}_{1\|}\simeq 0$ in smoothly varying inhomogeneous media. As a
result, in the framework of adiabatic approximation the transverse
mode decouples and this approximation renders polarization an
adiabatic invariant.

Expanding ${\bf E}_{1\bot}({\bf x})= \sum_{\lambda=\pm} {\bf
	e}_\lambda E_{1\bot\lambda}({\bf x}),$ makes the Eq. (6) an
eigenvalue equation
\begin{equation}
-\frac{1}{n^{2}(\bm{x},\omega)} \nabla^2
E_{1\bot\lambda}=\frac{\omega^2}{c^2}E_{1\bot\lambda},
\end{equation}
where ${\bf e}_\lambda$ are the orthonormal polarization vectors and
$\lambda =+ (-)$ stands for right (left)-handed polarization. The
differential operator in (7) is not Hermitian, However, by
introducing the Hermitian operator,
\begin{equation}
\hat{Q}=-\frac{1}{n({\bm x})}\nabla^2 \frac{1}{n({\bm x})},
\end{equation}
(7) can be written in the self-adjoint form
\begin{equation}
\hat{Q} \psi_\lambda=\frac{\omega^2}{c^2} \psi_\lambda,
\end{equation}
where
$$
\psi_\lambda({\bm x})=n({\bm x}) E_{1\bot\lambda}({\bm x}).
$$
Thus, with the weight function $n$, eigenfunctions
$E_{1\bot\lambda}$ form a complete orthogonal set. We demand a (slow
varying) periodic form for $n(\bm{x})$, so, according to Bloch
theorem, the eigenfunctions of (9) are Bloch functions, viz.,
\begin{equation}
\psi_{\lambda \textsf{n}} ({\bm x},{\bf q})= e^{i{\bf q}.{\bm x}}
u_{\lambda \textsf{n}} ({\bm x},{\bf q}), \label{Bloch}
\end{equation}
where ${\bf q}$ is the lattice momentum, $\textsf{n}$ is the band
index and $u_{\lambda \textsf{n}} ({\bm x},{\bf q})$ is periodic in
${\bm x}$ with the periodicity imposed on the plasma medium. The
Bloch waves (\ref{Bloch}) form a complete orthonormal set and, thus,
can be used as a convenient basis for constructing photons wave
packets. In view of our assumption of slow varying inhomogeneity and
in the spirit of adiabatic theorem, we drop the band index.

\section{The wave packet dynamics}

Let us proceed to the quantum level. The Bloch states are $\left|
{\psi _{\lambda }}({\bf q}) \right\rangle=e^{i{\bf q}.{\bm x}}
|u_{\lambda}({\bf q})\rangle$ and $\psi_{\lambda} ({\bm x},{\bf
	q})=\langle {\bm x} \left| {\psi _{\lambda }}({\bf q})
\right\rangle$ .We construct the (normalized) wave packet according
to
\begin{equation}
|\Psi(t) \rangle =\sum_\lambda z_\lambda(t) \int d^3{\bf q}\ a({\bf
	q},t) |\psi_\lambda({\bf q}) \rangle, \label{8}
\end{equation}
where $a({\bf q},t)$ and $z_\lambda(t)$ are amplitudes with the
normalizations
\begin{eqnarray}
\int {d^3{\bf q} | {a({\bf q},t)} |} ^2=1, \nonumber \\
\sum_\lambda  {| {z_\lambda(t)} |^2=1}. \nonumber
\end{eqnarray}
We assume that the momentum distribution $|a({\bf q},t)|^2$ is
sharply peaked around its mean value ${\bf q}_c$,
$$
{\bf q}_c(t)=\int d{\bf q}\ |a|^2{\bf q},
$$
which is the momentum of the wave packet. Writing the amplitude in
the form $a({\bf q},t)=\left| {a({\bf q},t)} \right|e^{-i\gamma
	({\bf q},t)}$, the wave packet center position, $\langle \Psi
|\hat{\bm x} | \Psi \rangle$, will be \cite{Torabi1,Sundaram}
\begin{equation}
{\bm x}_c =\int {d^3{\bf q}\left| a \right|^2 \{\nabla _{\bf q}
	\gamma ({\bf q},t)+} \sum\limits_{{\lambda }',\lambda }
{z^*_{{\lambda }'} z_\lambda \left\langle u_{{\lambda }'}({\bf q})
	\right|i\nabla _{{\bf q}} \left| u_{\lambda}({\bf q})
	\right\rangle}\}.
\end{equation}
According to our assumption of a well localized wave packet in the
momentum space, (12) reduces to
\[
{\bm x}_c(t)\approx {\bf \nabla }_{{\bf q}_c} \gamma({\bf
	q}_c,t)+\sum\limits_{{\lambda }',\lambda } {z^*_{{\lambda }'}
	z_\lambda {\bf A}_{{\lambda }'\lambda}({\bf q}_c)},
\]
where ${\bf A}_{{\lambda }'\lambda}({\bf q})=\left\langle
u_{{\lambda }'}({\bf q}) \right|i\nabla _{{\bf q}} \left|
u_{\lambda}({\bf q}) \right\rangle$. By introducing the notation
$|z(t)\rangle=\left( {{\begin{array}{*{20}c}
		{z_+(t) } \hfill \\
		{z_-(t) } \hfill \\
		\end{array} }} \right)$, the above equation will be
\begin{equation}
{\bm x}_c(t)\approx {\bf \nabla }_{{\bf q}_c} \gamma({\bf
	q}_c,t)+\langle z\mid\hat{{\bf A}}({\bf q}_c)\mid z\rangle,
\end{equation}
where $\hat{{\bf A}}$ is a $2\times2$ matrix with vector elements
${\bf A}_{{\lambda }'\lambda}$ and ${\bf A}_B=\langle z\mid\hat{{\bf
		A}}({\bf q}_c)\mid z\rangle$ is called Berry connection.

Let us derive equations of motion for such a wave packet. The wave
packet dynamics is obtained from the effective Lagrangian ($\hbar
=1$)
\begin{equation}
{\cal L}=i\langle \Psi | \dot{\Psi}\rangle-\langle \Psi |\hat{H}|
\Psi \rangle,
\end{equation}
using the Euler-Lagrange equation \cite{Jackiw}. The second term in
the Lagrangian is $\langle \Psi |\hat{H}| \Psi \rangle=\omega$, and
the first one will be
\begin{equation}
i\langle \Psi | \dot{\Psi}\rangle=\frac{\partial \gamma({\bf q}
	_c,t) }{\partial t}-i\langle z|\dot{z}\rangle.
\end{equation}
In deriving the above, we have used (11) and (13). By substituting
\[
\frac{\partial \gamma({\bf q} _c,t)}{\partial t}=\frac{d\gamma({\bf
		q} _c,t) }{dt}-{\dot {\bf q}}_c \cdot \nabla _{{\bf q}_c }
\gamma({\bf q} _c,t),
\]
in Eq. (15), the Lagrangian (14) takes the form
\begin{equation}
{\cal L}=i\langle z|\dot{z}\rangle+ {\dot {\bf q}}_c \cdot{\bf
	A}_B+{\bf q}_c \cdot {\dot {\bm x}}_c -\omega.
\end{equation}
We have omitted a term of total time derivative $d(\gamma -{\bm x}_c
\cdot {\bf q}_c)/dt$ in the Lagrangian, because it does not affect
the equations of motion. The equations of motion that follow from
the effective Lagrangian (16) are
\begin{equation}
{\dot {\bm x}}_{c} ={\bf v}_{g} + {\dot {\bf q}}_c\times {\bf
	\Omega}({\bf q}_c),
\end{equation}
\begin{equation}
{\dot {\bf q}}_c =\nabla_{{\bm x}_c} \omega,
\end{equation}
\begin{equation}
{|\dot z}_c\rangle=-i{\dot {\bf q}}_c\cdot  {\hat{\bf A}}({{\bf q}_c
}) |z_c\rangle,
\end{equation}
where ${\bf v}_g=\nabla_{{\bf q}_c} \omega$ is the group velocity
and ${\bf \Omega}=\nabla\times {\bf A}_B$ is called Berry curvature.

The second term on the right hand of Eq. (17), represents a
polarization dependent velocity perpendicular to the group velocity.
It is spin (polarization) Hall effect of photons. Eq. (19) describes
the time evolution of the polarization and has the solution
\begin{equation}
|z(t)\rangle=e^{i\int_C d^3{\bf q}_c\cdot {\hat{\bf
			A}}}|z(0)\rangle,
\end{equation}
where C is the wave packet trajectory in the parameter (momentum)
space. Eq. (20) can be written in the form
$|z(t)\rangle=e^{i\hat{\gamma}} |z(0)\rangle$, where
$\hat{\gamma}=\int_C d^3{\bf q}_c\cdot {\hat{\bf A}}$ is the
Berry-phase matrix. The phase shift is a geometric one which leads
to the rotation of the polarization plane for a linear polarized
wave known as the Rytov rotation law.

\section{Non-periodic inhomogeneous media}
The wave packet dynamics formalism is also applicable to the case of
slightly inhomogeneous {\it non-periodic} media. In this case the
basis for the wave packet expansion is $|\psi_\lambda({\bf q})
\rangle=e^{i{\bf q} \cdot {\bf x}} |{\bf e}_\lambda({\bf q})
\rangle$ where ${\bf e}_\lambda$ are the polarization vectors.
Therefore, we can replace $|u_\lambda\rangle$ by $|{\bf e}_\lambda
\rangle$ in the aforementioned formalism. We also write ${\bf
	e}_\lambda$ in the helicity basis as ${\bf e}_\lambda=
\frac{1}{\sqrt{2}} ({\bf i}_\theta \pm i {\bf i}_\phi)$, where ${\bf
	i}_\theta$ (${\bf i}_\phi$) is the zenithal (azimuthal) unit vector
in the spherical coordinates of the momentum space. Thence,
according to ${\bf A}_{{\lambda }'\lambda}({\bf q})=\left\langle
e_{{\lambda }'} \right|i\nabla _{{\bf q}} \left| e_{\lambda}
\right\rangle$, we will have
\[
\hat{\bf A}=\frac{\cot \theta}{q_c}\hat{\sigma}_3{\bf i}_\phi,
\]
where $\hat{\sigma}_3$ is the Pauli matrix. Thus, according to (20),
right/left circularly polarized waves experience a geometric phase
shift of $\pm \int_C \cos\theta\, d\varphi$, respectively.
$\theta(\varphi)$ is the zenithal (azimuthal) angle in spherical
coordinates. Therefore, if a linear polarized wave propagates in a
periodic plasma medium, the polarization plane will rotate which is
topological in nature. The Berry curvature will be
\[
{\bf \Omega } =\mp \frac{{\bf q}_c}{q_c^3},
\]
for right/left circularly polarized waves respectively, which,
according to (18), propagate along different trajectories and have
displacements perpendicular to the wave propagation direction. These
results of non-periodic inhomogeneous plasma media coincide with the
results of \cite{Torabi3} obtained from the post geometric
approximation.

\section{Conclusion}

We studied the propagation of electromagnetic waves in unmagnetized
slow varying periodic plasma media via the semiclassical wavepacket
approximation . This approximation enables us to solve problems in
the presence of periodic structures. The formalism yields to the
appearance of an Abelian gauge field (Berry connection) in the
effective lagrangian governing the wavepacket dynamics. This results
in the rotation of polarization plane, and the
polarization-dependent Hall effect for transverse photons in
periodic plasma media. The rotation of the polarization plane, of a
linear polarized wave, in unmagnetized periodic plasma media is
topological in nature and differs from the Faraday rotation which is
due to the presence of a magnetic field in the medium. The formalism
also applied to the case of non-periodic inhomogeneous plasma media.


\begin{thebibliography}{widest-label}
	
	\bibitem{Hacquin} S. Hacquin, S. Heuraux, M. Colin and G. Leclert, J. Comp. Phys. {\bf 174}, 1 (2001).
	\bibitem{Cho} S. Cho, Phys. Plasmas {\bf 11}, 4399 (2004).
	\bibitem{Malik} H. K. Malik, S. Kumer and K.P. Singh, Laser Part. Beams {\bf 26}, 197 (2008).
	\bibitem{Chen} F. F. Chen, \textit{Introduction to plasma physics and controlled fusion}, 2nd ed., Vol. 1 (Plenum Press, New York, 1984).
	\bibitem{Bakhmet eva} N. V. Bakhmet eva, V. V. Belikovich, M. N. Egereva, A. V. Tolmacheva, Radiophys. Quantum Electron., {\bf 53}, 69 (2010).
	\bibitem{Bliokh1} K. Yu. Bliokh and V. D. Freilikher, Phys. Rev. B {\bf 72}, 035108 (2005).
	\bibitem {Onoda1} M. Onoda, S. Murakami and N. Nagaosa, Phys. Rev. Lett. {\bf 93}, 083901 (2004).
	\bibitem{Onoda2} M. Onoda, S. Murakami and N. Nagaosa, Phys. Rev. E {\bf 74}, 066610 (2006).
	\bibitem{Torabi1} R. Torabi and M. Mehrafarin, JETP Lett. {\bf 88}, 590 (2008).
	\bibitem{Berry} M. V. Berry, Proc. R. Soc. A {\bf 392}, 45 (1984).
	\bibitem{Culcer} D. Culcer, A. MacDonald and Q. Niu, Phys. Rev. B {\bf 68}, 045327 (2003).
	\bibitem{Chang} M. C. Chang and Q. Niu, Phys. Rev. B {\bf 53}, 7010 (1996).
	\bibitem{Sundaram} G. Sundaram and Q. Niu, Phys. Rev. B {\bf 59}, 14915 (1999).
	\bibitem{Rytov} S. M. Rytov, Dokl. Akad. Nauk SSSR {\bf 18}, 263 (1938).
	\bibitem{Vladimirskii} V. V. Vladimirskii, Dokl. Akad. Nauk SSSR {\bf 13}, 222 (1941).
	\bibitem{Torabi2} R. Torabi, Can. J. Phys. {\bf 88}, 1 (2010).
	\bibitem{Dooghin} A. V. Dooghin, N. D. Kudnikova, V. S. Liberman and B. Ya. Zeldovich, Phys. Rev. A {\bf 45}, 8204 (1992).
	\bibitem{Liberman} V. S. Liberman, B. Ya. Zeldovich, Phys. Rev. A {\bf 46}, 5199 (1992).
	\bibitem{Torabi3} R. Torabi and M. Mehrafarin, JETP Lett. {\bf 95}, 277 (2012).
	\bibitem{Krall} N. A. Krall and A. W. Trivelpiece, Principles of Plasma Physics (McGraw-Hill, New York, 1973).
	\bibitem{Ginzburg} V. L. Ginzburg, \textit{The Propagation of Electromagnetic Waves in Plasmas} (Gordon and Breach, NY, 1964).
	\bibitem{Jackiw} R. Jackiw and A. Kerman, Phys. Lett. A {\bf 71}, 158 (1979).
	\bibitem{Kuo} S. P. Kuo, Progress In Electromagnetics Research B, {\bf 60}, 275 (2014).
	
	
	
\end{thebibliography}
\end{document}